\documentstyle[pra,aps]{revtex}
\begin{document}
\draft
\twocolumn[\hsize\textwidth\columnwidth\hsize\csname 
@twocolumnfalse\endcsname
\title{Propagation of the Angular Spectrum in the Up-conversion} 
\author{D. P. Caetano$^{*}$, M. P. Almeida and  P.H. Souto Ribeiro}
\address{Instituto de F\'{\i}sica, Universidade Federal do Rio de 
Janeiro, Caixa Postal 68528, Rio de Janeiro, RJ 21945-970, Brazil}
\author{A. Z. Khoury}
\address{Instituto de F\'{\i}sica,  Universidade Federal Fluminense, 
BR-24210-340 Niteroi, RJ, Brazil}
\author{C. H. Monken}
\address{Departamento de F\'{\i}sica, 
Universidade Federal de Minas Gerais, Caixa Postal 702, 
Belo Horizonte, MG 30123-970, Brazil} 
\date{\today}
\maketitle
\begin{abstract}
We study theoretical and experimentally the propagation of the angular spectrum
of the light produced in the up-conversion process.
The connection between the angular spectrum of the fundamental and the
second harmonic is derived and measured. We show that even though they
are connected, it is not possible to directly transfer images from the fundamental
to the second harmonic.
\end{abstract}
]

\section{Introduction}

Parametric up and down-conversion has been subject of investigation over the past
few decades\cite{1a,1b}. The special case of collinear up-conversion, known as
second harmonic generation(SHG) has received a great deal of attention, because it has
become a reliable method for generating laser beams in a vast range of
wavelengths from the green region up to ultra-violet. In both cases, up and
down-conversion, the transverse properties of the field were initially neglected. 
The interest in the transverse properties of the fields involved in the non-linear
processes is increasing over the past few years\cite{2}.
In most of the theoretical and experimental investigations, the fields involved are 
represented by Hermite-Gaussian modes. More recently, the 
interest in optical modes possessing orbital angular momentum, described by Laguerre-Gaussian modes, has found in
SHG, a way of obtaining modes with higher orbital angular momentum, as it
is doubled in this process\cite{3}. 

In parametric up and down-conversion processes, the converted fields are
connected to the pumping fields through phase matching conditions. It has
been shown, for example, that the entangled field of twin photons from 
down-conversion carries the same angular spectrum as the pump\cite{4}. It was
also shown that in stimulated down-conversion, the idler beam may carry the same
angular spectrum as the pump or the complex conjugated of the stimulating beam\cite{5}.
This transfer of angular spectrum implies in transfer of images formed
by those beams and also in transfer of orbital angular momentum\cite{3,6,7}.

The propagation of the transverse properties of 
a light field,  like orbital angular momentum and many others, are defined by the propagation
of its angular spectrum. In this work, we study the propagation of light
fields and their coupling in the SHG process, taking into account their transverse 
spatial distributions, or in other words, propagating their angular spectra.
We demonstrated how the angular spectrum of the fundamental and the second harmonic
field are connected. It is shown that even though they are
connected, this connection does not allow direct transfer of images from one 
field to the other. On the other hand,  as the transfer function is known, it is possible 
to manipulate the second harmonic angular spectrum through preparation of the fundamental.

\section{Theory}

Let us consider the experimental set-up sketched in Fig.\ref{fig1}. In the noncollinear case (Fig.\ref{fig1}a) two different modes pump a non-linear crystal for generating a third mode, the up-converted one. SHG is the collinear version (Fig.\ref{fig1}b), where just one mode pumps the non-linear crystal and the degenerate collinear up-conversion occurs.

In the following, we present a quantum theory for the up-conversion,
based on the formalism developed by Mandel and co-workers\cite{1a}.
We assume that depletion of the fundamental field is negligible in the parametric interaction, the second harmonic field is weak and the monochromatic and paraxial approximations are used. The calculation of the effects of propagation and transfer of the angular spectrum in 
up-conversion does not require a quantum theory. The same results
would be obtained with a classical formalism, however the use of the
quantum hamiltonian and the deduction of the quantum state for the
up-converted field may be helpfull in future developements. 

The quantum state of the up-converted field is obtained in the same
way as it is done in Ref.\cite{5} for spontaneous and stimulated
down-conversion. The difference is that here, signal and idler fields
are in coherent states while the up-converted field is a multimode Fock state with n=1: 

\begin{equation}
\label{eq1}
|\psi\rangle = |vac\rangle + C\int d\bbox{q}_2 \int \,\, d\bbox{q}_3 \,\, v_1(\bbox{q}_3-\bbox{q}_2)
 v_2(\bbox{q}_2) |1 ; \bbox{q}_3\rangle,
\end{equation}
where $v_1(\bbox{q}_3-\bbox{q}_2)$ and  $v_2(\bbox{q}_2)$ are the angular spectra of the input 
modes with transverse component of the momentum $\bbox{q}_3-\bbox{q}_2$ and $\bbox{q}_2$, repectively, and $|1;\bbox{q}_3\rangle$ is a Fock state with n=1 for the up-converted mode with the 
transverse component of the momentum $\bbox{q}_3$.

The transverse intensity profile of the up-converted field in a plane
situated at a distance $z$ from the crystal is given by:

\begin{eqnarray}
I(\bbox{r})=&&\langle E^{(-)}(\bbox{\rho},z)E^{(+)}(\bbox{\rho},z)\rangle,
\label{eq4}
\end{eqnarray}
where the electric field operator is written as:

\begin{eqnarray}
E^{(+)}(\bbox{\rho},z)= \int \,\, d\bbox{q}\prime a(\bbox{q}\prime)
\exp \biggr[ i\left(\bbox{q}\prime\cdot\bbox{\rho} -
\frac{q\prime^{2}}{2k\prime}z\right)\biggr].
\label{eq3}
\end{eqnarray}

Performing the calculation we obtain:

\begin{eqnarray}
\label{eq5}
I(\bbox{r}) \propto  \biggr | \int \,\, d\bbox{\rho}\prime  \,\,
{\cal W}_1(\bbox{\rho}\prime) {\cal W}_2(\bbox{\rho}\prime)
\exp \left[ i |\bbox{\rho} - \bbox{\rho}\prime|^{2}\frac{k}{2z}\right]
\biggr|^{2} .
\end{eqnarray}

For the special case where modes 1 and 2 are the same (collinear case),
the above equation is simplified to

\begin{eqnarray}
\label{eq6}
I(\bbox{r}) \propto  \biggr | \int \,\, d\bbox{\rho}\prime \,\,
{\cal W}^2(\bbox{\rho}\prime) 
\exp \left[ i |\bbox{\rho} - \bbox{\rho}\prime|^{2}\frac{k}{2z}\right]
\biggr|^{2} .
\end{eqnarray}

From the equations above it is seen that the transverse field of the second harmonic
is given by the square of the fundamental, propagated from the crystal to the
observation  plane.

\subsection{Propagation of a double-slit pattern}

Let us consider a plane wave as input to a double-slit, as it is represented in
Fig.\ref{fig2}. Even though this is a two-dimension problem, we are going to calculate 
the transverse field distribution at the crystal plane for one dimension, for simplicity.
After the slits, the beam is focused by a thin lens inside the crystal. The lens is 
necessary because the efficiency of the up-conversion is low and focusing improves the 
efficiency\cite{8}.
The outcome of this calculation will be used in Eq.\ref{eq6} for calculating the final
intensity distribution of the SHG field. 
At the plane of the slits, the field is given by:

\begin{equation}
\label{eq7}
{\cal W}(x,0) = S(x+\frac{d}{2}, a) + S(x-\frac{d}{2}, a),
\end{equation}
where

\begin{equation}
\label{eq7a}
S(x,a) = \left\{
\begin{array}{ll}
 1 &  \mbox{if $|x| \le \frac{a}{2}$} \\
 0 &  \mbox{elsewhere},
\end{array}
\right.
\end{equation}
$a$ is the slit width and $d$ is the distance between them.
The angular spectrum is calculated:

\begin{eqnarray}
\label{eq8}
v(q , 0) \propto &  \int dx \,\, {\cal W}(x, 0) \exp (-i q x) \\ \nonumber
                   \propto &  \mbox{sinc}(q\frac{a}{2}) \cos(q\frac{d}{2}),
\end{eqnarray}
where $q$ is the projection of the transverse wavevector $\bbox{q}$ in the
$x$ direction.
The angular spectrum is easily propagated to the plane just before the lens:

\begin{equation}
\label{eq9}
v^-(q , z_0) \propto \mbox{sinc}\left(q\frac{a}{2}\right) \, \cos\left(q\frac{d}{2}\right)
\,\, exp \left(-i \frac{q^2}{2k}z_0\right),
\end{equation}
where the paraxial approximation was taken and $k$ is the wavenumber.
In the plane just after the lens, the angular spectrum is given by:

\begin{eqnarray}
\label{eq10}
v^+(q , z_0) \propto \int \,\, d\xi \,\, \mbox{sinc}\left( q\frac{a}{2}\right) \,\, 
\cos\left( q\frac{d}{2} \right) \times \\ \nonumber
\times \exp \left[ -i \frac{(z_0 - f)}{2k}\xi^2\right]   \exp \left(-i \frac{f}{k}q\xi\right).
\end{eqnarray}

After propagation to the plane inside the crystal and turning it back into a field distribution, 
we obtain : 
\begin{eqnarray}
\label{eq11}
&{\cal W}(x, z_0 + f) \propto \mbox{sinc}\left(\frac{k}{f}\frac{a}{2} x\right) 
\cos\left(\frac{k}{f}\frac{d}{2} x\right) \times \\ \nonumber
& \times \exp\left[-i \frac{k}{2f^2}(z_0 - f) x^2\right].
\end{eqnarray}

From the field in Eq. \ref{eq11} above, one can calculate the final transverse distribution for both fundamental and second harmonic fields, in a plane situated at an arbitrary
distance $z$ from the crystal. For the fundamental it is just a matter of propagating the angular
spectrum again and for the second harmonic one will make use of Eq.\ref{eq6}.

\subsection{Image plane}

The slits are imaged by the lens in a plane after the crystal. The longitudinal
position of this plane is given by the usual relation 

\begin{equation}
\label{eq14}
\frac{1}{f} = \frac{1}{i} + \frac{1}{o},
\end{equation}
where $o$ is the distance between the slits(object) and the lens and $i$ is the
distance between the lens and the image. According to our scheme, decribed in 
Fig.\ref{fig2}, $o = z_0$ and $i = f + z_D$.

In order to obtain the field distributions at the image plane, we write
the propagation integral for fundamental and second harmonic, starting from
Eq.\ref{eq11}, in terms of the real parts of the fields:

\begin{eqnarray}
\label{eq15}
&{\cal W}(x,z_0+f+z_D) \propto \int d\xi \,\, {\cal W}_R(\xi, z_0 + f) \times \\ \nonumber
&\times \exp\left\{ ik \left[\left(\frac{1}{2z_D} - \frac{z_0}{2f^2} + \frac{1}{2f}\right)\xi^2 - 
\frac{1}{z_D}\xi x + \frac{x^2}{2z_D}\right]\right\},
\\ \nonumber & \mbox{for the fundamental and} \\ \nonumber
&{\cal T}(x,z_0+f+z_D) \propto \int d\xi \,\, {\cal W}_R^2(\xi, z_0 + f) \times \\ \nonumber
&\times \exp\left\{ 2ik \left[\left(\frac{1}{2z_D} - \frac{z_0}{2f^2} + \frac{1}{2f}\right)\xi^2 - 
\frac{1}{z_D}\xi x + \frac{x^2}{2z_D}\right]\right\},
\\ \nonumber & \mbox{for the second harmonic.}
\end{eqnarray}

The image is obtained when the quadratic part of the propagation phase is zero:

\begin{equation}
\label{eq16}
\frac{1}{2z_D} - \frac{z_0}{2f^2} + \frac{1}{2f} = 0.
\end{equation}

This condition is achieved in both cases, when Eq.\ref{eq14} is satisfied. 
For the fundamental, it is easy to see that when the quadratic term is zero,
the field is given by the Fourier transform of ${\cal W}_R(\xi, z_0+f)$:

\begin{eqnarray}
\label{eq17}
&{\cal W}(x,z_0+f+z_D) \propto  \\ \nonumber
& \propto \int d\xi \,\, {\cal W}_R(\xi, z_0 + f) 
\exp\left( ik \frac{1}{z_D}\xi x \right)  \\ \nonumber
 & \propto  S\left[\left( x+\frac{d}{2}\right) \frac{f}{z_D},a\frac{f}{z}\right] +
 S\left[\left( x-\frac{d}{2}\right) \frac{f}{z_D},a\frac{f}{z_D}\right],
\end{eqnarray}
the  image of the slits is amplified by the factor $\frac{z_D}{f}$.
For the second harmonic however, we have:

\begin{eqnarray}
\label{eq18}
&{\cal T}(x,z_0+f+z_D) \propto \\ \nonumber
& \propto \int d\xi \,\, {\cal W}^2_R(\xi, z_0 + f) 
\exp\left( ik \frac{1}{z_D}\xi x \right).
\end{eqnarray}

The solution of Eq.\ref{eq18} is the self-convolution  of the Fourier transform
of the function ${\cal W}_R(\xi, z_0 + f)$, which is the double-slit function
defined in Eq.\ref{eq17}. It is plotted in Fig.\ref{fig3} considering the parameters
of our set-up.
As we can see, the image formed by the fundamental is not formed by the
second harmonic.

\subsection{Far field}

The intensity distributions for the fundamental and the second harmonic
in a plane situated at a distance $z_D$ from the crystal, which is after
the image plane, can be obtained from the fundamental field inside the crystal, 
given by Eq.\ref{eq11}. The analytical solution for the evolution integrals is not
straightfoward in this case. We have chosen to solve this problem by replacing the real part of the function in Eq.\ref{eq11} with a sum of gaussians, as an approximation to solve the propagation integrals. Although a closed form of the solution is obtained, resulting function is not instructive for the reader. We present instead, in Fig.\ref{fig4}, a plot this function, for the parameters of our set-up, in order to compare it with the experimental results.

\section{Experiment}
We have performed measurements of the transverse intensity distribution for
the fundamental and the second harmonic in the SHG, in different propagation
planes after the crystal.

An infrared diode laser, with wavelength centered around 845 nm, 150 mW output
power is directed to a LiIO$_3$ (Lithium Iodate) non-linear cristal. SHG takes
place, producing a beam with wavelength around 425 nm. Before the infrared reaches
the crystal, it is passed through a double-slit diffraction screen with slits width
$a$ = 0.2mm, separation $d$ = 0.4mm and also through a
thin lens with $f$ = 10 cm focal length. After the crystal, fundamental and second
harmonic beams are separated by a prism and they are both detected with single
photon counting modules, based on cooled avalanche photodiodes. The photon counting
is necessary for the second harmonic beam, because the signal level is 
strongly decreased by the passage of the fundamental through the double-slit.
The pulses coming from the detectors are sent to photon counters controlled by
a computer which performs the data acquisition.

\section{Results and discussion}

In a first set of measurements we have placed the slits at 12.3cm before the lens,
so that the image of the slits were formed in a plane situated at 53.4cm after the lens.
The crystal is always in the focal plane of the lens, so that the detection plane is
43.4cm from its center. The detectors are scanned in the vertical direction and the
intensity profile is registered for both the fundamental and the second harmonic
signals. Fig.\ref{fig5} shows these profiles. The fundamental presents an extended
image of the slits, and the amplification factor is in good agreement with Eq.\ref{eq17},
since $d \simeq$ 0.2 mm, $\frac{z_D}{f} \simeq$ 4 and the distance between slits
in the amplified image is $d\prime \simeq$ 0.8 mm.
The second harmonic is also in agreement with Eq.\ref{eq18}, which is plotted in Fig.\ref{fig3}.
The  measurement procedure with a finite detection slit smooths the curve. 
This result shows that images can not be directly transferred from the fundamental 
to the second harmonic. On the other hand, the basic
information about the image is contained in the second harmonic pattern and it might
be recovered by deconvolution.

In order to check the evolution of the fields, starting from the crystal as it was
calculated, we have measured the intensity distributions inside the crystal. These
measurements were performed placing a second lens after the crystal, imaging the
center of the crystal onto the detection plane. The results are displayed in 
Fig.\ref{fig6}. These patterns are described by Eq.\ref{eq11}, where the
intensity of the fundamental is given by the square modulus of the field and the
intensity of the second harmonic is given by the square modulus of the square of the
field. 

In a second set of measurements, the slits were placed 2cm before the crystal
so that no real image was formed for the fundamental. In this case, the far field
calculation must be used. The results are shown in Fig.\ref{fig7}a for the 
fundamental and Fig.\ref{fig7}b for the second harmonic. 
As before, we have also measured the intensity distributions inside the crystal.
The results are shown in Fig.\ref{fig8}. These curves should be fitted to the
square of the field in Eq.\ref{eq11}. However, we have used gaussian functions for 
fitting the patterns in Fig.\ref{fig8}. The parameters obtained with the nonlinear fittings were used in the calculations for generating curves in Fig.\ref{fig4}. Comparing the intensity profiles in Fig.\ref{fig4} and the measured distributions in Fig.\ref{fig7}, a good agreement is obtained.

For the fundamental, the far field intensity pattern is just the same as inside the
crystal (Fig.\ref{fig8}a), enlarged by a scale factor due to the propagation(Fig.\ref{fig7}a). 
For the second harmonic, the pattern outside the crystal is different. The oscillations
present two spatial frequencies. As a matter of fact, we have fitted the experimental
data in Fig.\ref{fig7}b to a two-frequency interference pattern:

\begin{eqnarray}
\label{eq20}
&I(x, z_0 + f + z_D) \propto \mbox{sinc}^4\left(\frac{k}{f}\frac{a}{2} x\right) \times \\ \nonumber
&\times \left[ 1 + \mu_1 \cos\left(2\frac{k}{f}\frac{d}{2} x\right) + 
\mu_2 \cos\left(4\frac{k}{f}\frac{d}{2} x\right)\right] .
\end{eqnarray}

The same function could be used to fit the intensity pattern of the second harmonic
inside the crystal, Fig.\ref{fig8}b. The reason why Fig.\ref{fig7}b and Fig.\ref{fig8}b
look different is that they have different visibilities $\mu_1$ and $\mu_2$. During the propagation the ratio between these visibilities changes as it is shown in the theoretical result of Fig.\ref{fig4}b.

\section{Conclusion}

In conclusion, we have derived the transverse intensity distribution for the light beam produced in the up-conversion process, as a function of the input pumping fields.
We have also calculate the intensity distributions for the input fundamental and the output
second harmonic beam generated in the SHG process, when the fundamental is diffracted
through a double-slit before the non-linear crystal. The distributions are calculated for
the propagation planes inside the crystal, the image plane and the far field.
It is demonstrated experimentally, that images formed on the fundamental are not
transferred to the second harmonic. It is also demonstrated that the double-slit
interference pattern transferred to the second harmonic, oscillates with two spatial
frequencies. The experimental data are in good agreement with the calculation.

Financial support was provided by Brazilian agencies CNPq, PRONEX, FAPERJ, 
FUJB and Institutos do Mil\^enio-Informa\c c\~ao Qu\^antica.

\begin{figure}[h]
%\vspace*{5cm}
%\special{ps:fig1a.ps x=5cm y=6cm}
%\vspace*{6cm}
%\special{ps:fig1b.ps x=5cm y=6cm}
\caption{Sketch of the experiment. a) Non-collinear up-conversion. 
b) Second harmonic generation.}
\label{fig1}
\end{figure}

\begin{figure}[h]
%\vspace*{4cm}
%\special{ps:fig2.ps x=8cm y=4cm}
\caption{Propagation through a double-slit and a thin lens.}
\label{fig2}
\end{figure}

\begin{figure}[h]
%\vspace*{4cm}
%\special{ps:fig3.ps x=5cm y=4cm}
\caption{Self-convolution for a two-slits distribution.}
\label{fig3}
\end{figure}

\begin{figure}[h]
%\vspace*{4cm}
%\special{ps:fig4.ps x=8cm y=4cm}
\caption{Theoretical intensity patterns. a) Fundamental, far field, $z = z_0 + f + z_D$. 
b) Second harmonic, far field, $z = z_0 + f  + z_D$.}
\label{fig4}
\end{figure}

\begin{figure}[h]
%\vspace*{4cm}
%\special{ps:fig5.ps x=8cm y=4cm}
\caption{Experimental intensity patterns. a) Fundamental, image plane, $z = z_0 + f + z_D$. b) Second harmonic, image plane, $z = z_0 + f + z_D$.}
\label{fig5}
\end{figure}

\begin{figure}[h]
%\vspace*{4cm}
%\special{ps:fig6.ps x=8cm y=4cm}
\caption{Experimental intensity patterns. a) Fundamental, crystal plane, $z = z_0 + f $. 
b) Second harmonic, crystal plane, $z = z_0 + f $.}
\label{fig6}
\end{figure}

\begin{figure}[h]
%\vspace*{4cm}
%\special{ps:fig7.ps x=8cm y=4cm}
\caption{Experimental intensity patterns. a) Fundamental, far field, $z = z_0 + f + z_D$. 
b) Second harmonic, far field, $z = z_0 + f  + z_D$.}
\label{fig7}
\end{figure}

\begin{figure}[h]
%\vspace*{4cm}
%\special{ps:fig8.ps x=8cm y=4cm}
\caption{Experimental intensity patterns. a) Fundamental, crystal plane, $z = z_0 + f $. 
b) Second harmonic, crystal plane, $z = z_0 + f $.}
\label{fig8}
\end{figure}

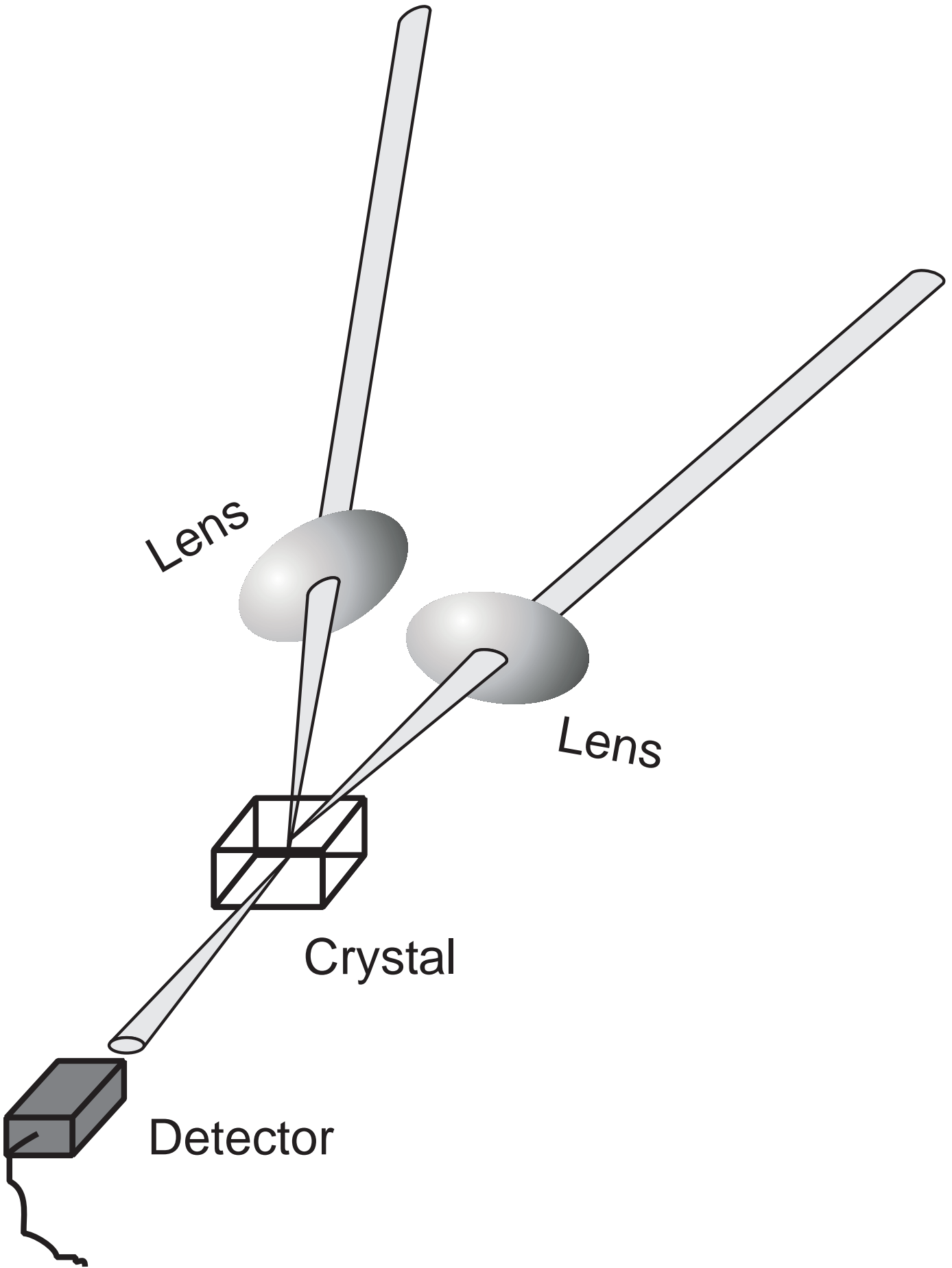
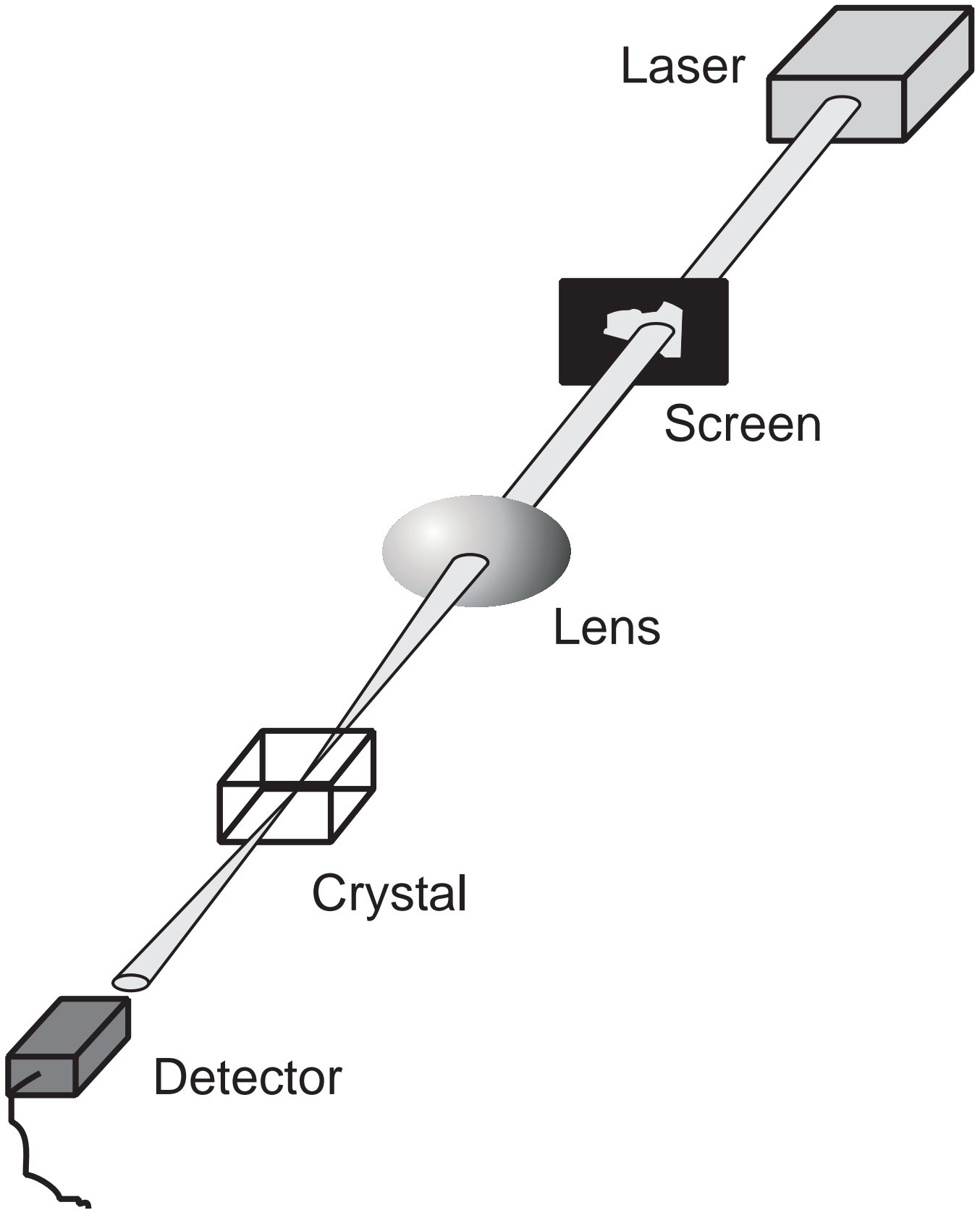
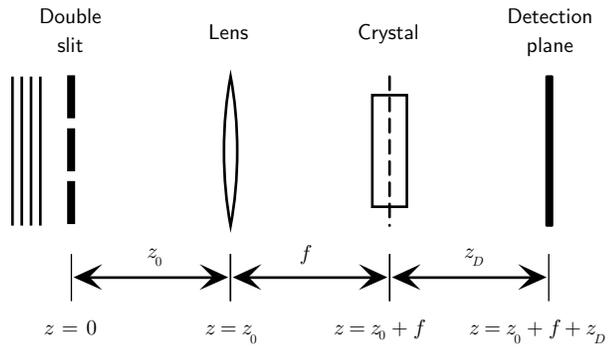
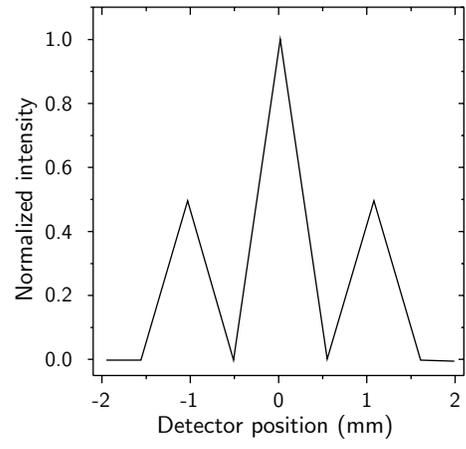
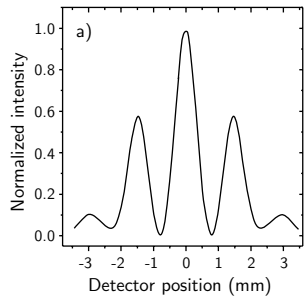
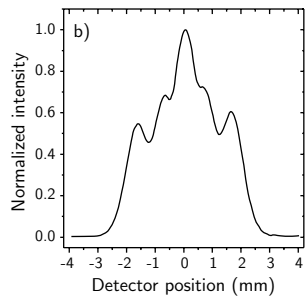
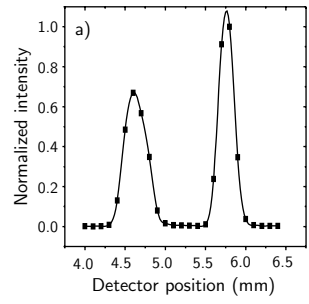
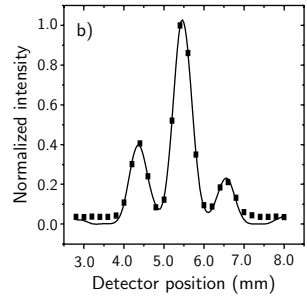
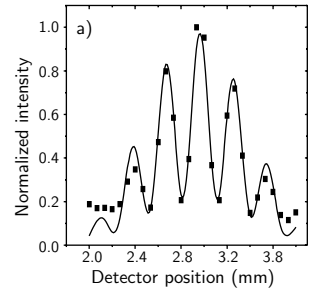
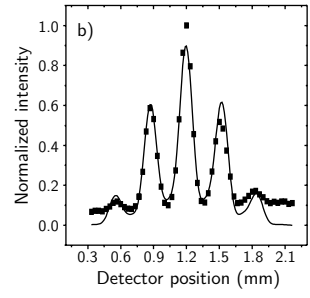
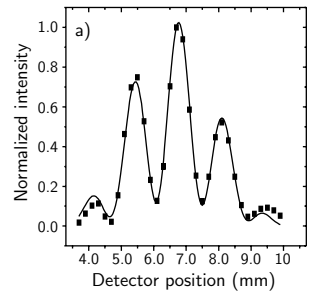
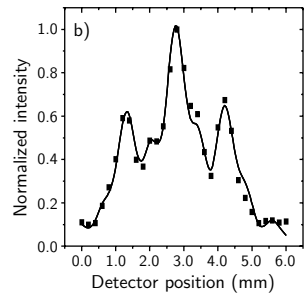
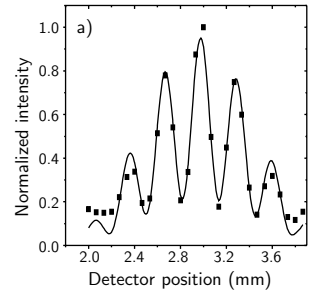
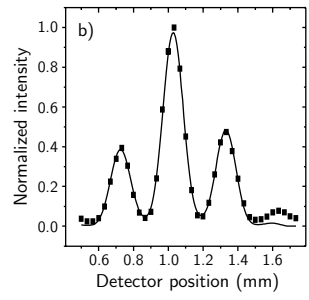
\end{document}